# Double-Helix Singularity and Vortex-Antivortex Annihilation in Space-Time Helical Pulses

Shuai Shi[1], Ren Wang[1,2,*], Minhui Xiong[1], Qinyu Zhou[1], Bing-Zhong Wang[1], Yijie Shen[3,4,*]

1 Institute of Applied Physics, University of Electronic Science and Technology of China, Chengdu, China

2 Yangtze Delta Region Institute (Huzhou), University of Electronic Science and Technology of China, Huzhou, China

3 Centre for Disruptive Photonic Technologies, School of Physical and Mathematical Sciences & The Photonics Institute, Nanyang Technological University, Singapore, Singapore

4 School of Electrical and Electronic Engineering, Nanyang Technological University, Singapore, Singapore

*Corresponding author: Ren Wang (rwang@uestc.edu.cn) and Yijie Shen (yijie.shen@ntu.edu.sg)

**Abstract:** Topological structures reveal the hidden secrets and beauty in nature, such as the double helix in DNA, whilst the manipulation of which in physical fields, especially in ultrafast structured light, draw booming attention. Here we introduce a new family of spatiotemporal light fields, i.e. helical pulses, carrying sophisticated double-helix singularities in its electromagnetic topological structures. The helical pulses were solved from Maxwell's equation as chiral extensions of toroidal light pulses but with controlled angular momentum dependence. We unveil that the double helix singularities can maintain their topological invariance during propagation and the field exhibits paired generation and annihilation of vortices and antivortices in ultrafast space-time, so as to be potential information carriers beating previous conventional vortex structured light.

## 1 Introduction

The double helix, an intriguing structure composed of two intertwined helical strands, is widely recognized as a chiral topological structure, most famously exemplified by deoxyribonucleic acid (DNA) [1-3]. Double-helix structures have also been observed in two-phase dendrites [4], macromolecular phases [5], superconductors [6], supercrystals [7], and topological magnetic fields [8].

Inspired by structures with unique topologies, structured light has recently garnered widespread attention and research interest, such as diverse OAM beams [9-11], Möbius strips [12, 13], knots [14-16], C point V points L lines in Poincare beams [17-19], optical skyrmions [20-24], and optical hopfions [25-27]. As an important part of the topology structure, the singularity is a significant characteristic of structured lights, the structure of the singularity indicates the flow direction of the field and reveals the topological structure of the field from another perspective, such as line-type phase singularities [28, 29] in vortex beams [30], and point-type polarization singularities [31-33] in vector beams [34, 35]. The topologies with kinds of phase and polarization singularities have opened up new opportunities for the application, such as light-matter interaction [36, 37], nonlinear optics [38], quantum processing [39, 40], microscopy and imaging [41], metrology and information transmission [42-44], etc. Recently, a famous spatiotemporal toroidal light pulses [45] proposed by Hellwarth and Nouchi [46] named "Flying Doughnut" was experimentally observed in optical [47], THz [48] and microwave [49] frequency range, which has space-time nonseparability [50], skyrmion topologies [51], and can be couple to anapole [52, 53]. Moreover, toroidal light pulses display the singularities including saddle points and vortices and nulls including spherical shells and rings [54]. However, no reports of double helix structures within structured light have been documented.

In this paper, we introduce a new family of structured light and discover double helix singularities within these light pulses, closely resembling the structure of DNA. This type of singularity exists in the electric field of helical pulse [55] with azimuth dependence, which belong to an extended family of toroidal pulses. Moreover, we reveal the existence of paired generation and annihilation of vortices and antivortices in helical pulses. The annihilation phenomenon is reported in stable propagating topologic light quasi-particle for the first time.

## 2 Theory and calculation

Following the electromagnetic directed-energy pulse trains derivation method [56], to obtain the electric and magnetic fields for helical pulses, we start with an appropriate scalar generating function $f(\mathbf{r},t)$ that satisfies Helmholtz's wave equation $\left(\nabla^2 - \frac{1}{c^2}\frac{\partial^2}{\partial t^2}\right)f(\mathbf{r},t)=0$, where $c=1/\sqrt{\mu_0\varepsilon_0}$, is the speed of light, and 4 $\varepsilon_0$ and $\mu_0$ are the permittivity and permeability of the medium, respectively. Next, the exact solution of $f(\mathbf{r},t)$ can be given by the combination of modified power spectrum method and helical phase factor [55], as

$$f(\mathbf{r},t) = \left(\frac{\rho}{q_1 + i\tau}\right)^{|l|} e^{il\theta} \frac{f_0}{\rho^2 + (q_1 + i\tau)(q_2 - i\sigma)}, \quad \text{Where}$$

$\tau = z - ct$, $\sigma = z + ct$, $f_0$ is a normalizing constant and $l$ is a constant defining the topological number or the order due to the spatial continuity of electromagnetic fields, we limit $l$ to an integer. When compared to a Gaussian beam, the parameters $q_2$ and $q_1$ represent respectively the "Rayleigh range" (or depth of the focal region) and effective wavelength [46]. In particular, the value of the ratio $q_2/q_1$ indicates whether the pulse is strongly focused or collimated ($q_2/q_1 \gg 1$), in the content studied in this article, $q_1$ and $q_2$ will affect the intercept and radius of the singular vortex line. The electromagnetic fields for the transverse electric (TE) solution can be derived by the potential $\Pi = \nabla \times \hat{z} f(\mathbf{r},t)$ as $E(\mathbf{r},t) = -\mu_0 \frac{\partial}{\partial t} \nabla \times \Pi$ and $H(\mathbf{r},t) = \nabla \times (\nabla \times \Pi)$. Finally, in a cylindrical coordinate system, the TE field components are given by the expressions:

$$E_\rho = \frac{f_0 l e^{il\theta}}{\rho} \sqrt{\frac{\mu_0}{\varepsilon_0}} \left(\frac{\rho}{q_1+i\tau}\right)^{|l|} \left\{ \frac{(q_2+q_1-2ict)}{\left[\rho^2+(q_1+i\tau)(q_2-i\sigma)\right]^2} + \frac{|l|}{\left[\rho^2+(q_1+i\tau)(q_2-i\sigma)\right](q_1+i\tau)} \right\} \quad (1)$$

$$E_\theta = if_0 \sqrt{\frac{\mu_0}{\varepsilon_0}} e^{il\theta} \left(\frac{\rho}{q_1+i\tau}\right)^{|l|} \left\{ \frac{l^2}{\left[\rho^2+(q_1+i\tau)(q_2-i\sigma)\right]\rho(q_1+i\tau)} + \frac{|l|(q_2+q_1-2ict)}{\rho\left[\rho^2+(q_1+i\tau)(q_2-i\sigma)\right]^2} - \frac{4\rho(q_2+q_1-2ict)}{\left[\rho^2+(q_1+i\tau)(q_2-i\sigma)\right]^3} - \frac{2|l|\rho}{\left[\rho^2+(q_1+i\tau)(q_2-i\sigma)\right]^2(q_1+i\tau)} \right\} \quad (2)$$

where $E_\rho$ and $E_\theta$ represent the radially and azimuthally directed electric field component, the magnetic field can be determined based on Maxwell equation. Note that the transverse electric (TE) mode field does not possess longitudinally directed components of electric field $E_z$. The transverse magnetic (TM) mode can be expressed by exchanging the electric and magnetic fields. For $l=0$, the electromagnetic fields in Eq. (S8) are reduced to the fundamental toroidal pulse [46], which has only one field components. Due to limited energy constraints, it is necessary to limit $|l| \leq 2$, the proof process is included in Supplementary Note 1, this article only discussing the case where $l=1$. Moreover, the real and imaginary parts of Eq. (S8) simultaneously satisfy Maxwell equations, and it can be verified that the imaginary electric field can be obtained by rotating the real electric field $\pi/2l$. See detailed derivation of Eq. (S8) in Supplementary Note 1.

The spatial topology of the electric field and the singularity lines of a TE helical pulse with $q_2=10q_1$ and $q_1=0.01$ near the focus ($t=0$) are shown in Figure 1. Figure 1 (a) - (c) reveal in detail the dynamic evolution of $E_\rho$, $E_\theta$ and $|E|$ of the TE helical pulse electric field during its propagation process, due to the Gouy phase shift, the pulse transformation between the $1\frac{1}{2}$-cycle and single-cycle. In Figure 1 (c), two intertwined curve are presented at each moment, where the red curve represents the path connected by a right-handed vortex core, while the blue curve corresponds to the connection path of a left-handed vortex core. These curves not only cross the main part of the wave packet at every moment, but also continue to extend towards infinity after completing a mutual entanglement within the main area. It is worth noting that throughout the entire propagation process, the singularity structures maintain topological invariance and continue to exhibit a double helix geometric shape. Figure 1 (d) shows schematic diagram of Double-helix Singularities. Red and blue represent the singularity lines, while yellow and green arrows indicate the direction of the electric field vectors. At the same time, figure 1 (e) shows the structure of DNA. The similarity of the double helix structure is manifested.

## 3 Results

### 3.1 Double-helix Singularities

We employ method for determining singularities in nonlinear differential equation system to determine the nature of the two vortices, similar to the theory of Helical vortices in swirl flow mechanics. the specific principles and results of this determination are provided in the Supplementary Note 4. Figure 2(a) shows the electric field vector on several planes and singularity line at focus ($t=0$). Figure 2(b)-(d) shows the electric field vector on $z=2q_1$, $q_1$, 0, the red and blue arrows indicate the direction of the electric field. The electric field vector forms two vortices with opposite chiralities around the singularity on each transverse plane. We use red to mark the vortex core of the right-handed vortex field and blue to mark the vortex core of the left-handed vortex field. On $z=2q_1$ plane, the two vortex cores are located in the regions $y>0$ and $y<0$, respectively. During the rise to the $z=0$ plane, the two vortex cores complete a 180-degree rotation, exchange their regions, and drive the singularity line to achieve a half-cycle spiral winding. Continuing to rise, the two vortex cores complete another position exchange, eventually returning to their original regions, and finally driving the singularity line to achieve a complete helix. During the propagation time $t \in (-q_2/c, q_2/c)$, the helical singularity structure propagates with the pulse, remaining the helical topology. Figure 3(a) shows the singularity line image of the central region of the wave packet in Figure 3(b), showing the topological invariance of the double helix structure during propagation. Please refer to the supplementary video for the evolution of the electric field singularity.

### 3.2 Annihilation of Vortex and Antivortex

Unlike the stage $t \in (-q_2/c, q_2/c)$, the singularity lines undergo two bending events when the motion occurs in $t \in (-\infty, -q_2/c)$ and $t \in (q_2/c, \infty)$ stage. Figure 3(b) illustrates the singularity points and fields on different planes at $t=-15q_1/c$. Figures 3(c1)-(c4) present the electric field vector flow maps in the bending region and at the two critical planes, respectively. It can be clearly observed that there are six singularities present on these planes, including two antivortices and four vortices. We have identified and characterized the vortices and antivortices, and the specific process and results are provided in the Supplementary Note 4. To investigate the birth and annihilation processes of these singularities, we labeled the six singularities and plotted vector field diagrams in the vicinity of two critical planes. On the $z=25q_1$ plane, vortex No.1 and antivortex No.2 are generated as a pair, and vortex No.5 and antivortex No.6 are also generated as a pair. During the plane $z=25q_1$ and $z=40q_1$, the antivortex No. 2(5) and vortex No. 3(4) move closer to each other, while vortex No. 1 and vortex No. 6 move away from the central area and continue to move outward. On the $z=40q_1$ plane, antivortex No.2 annihilates with vortex No.3, and vortex No.4 annihilates with vortex No.5. On this plane, only vortex No.1 and vortex No.6 remain in the space, and they maintain the property of vortex dipole.

Overall, apart from the bending region, the conditions in other areas are identical to those in $t \in (-q_2/c, q_2/c)$. On any plane outside the bending segment, there exist two vortices with opposite rotational directions. In the central region of the wave packet, two singularity lines approach each other, twist around once, and eventually separate. The above content shows the situation at a time point during the $t \in (-\infty, -q_2/c)$ stage, where the bending state occurs in the $z > ct$ region. In the $t \in (q_2/c, \infty)$ stage, the situation is exactly the opposite, and the bending state occurs in the $z < ct$ region.

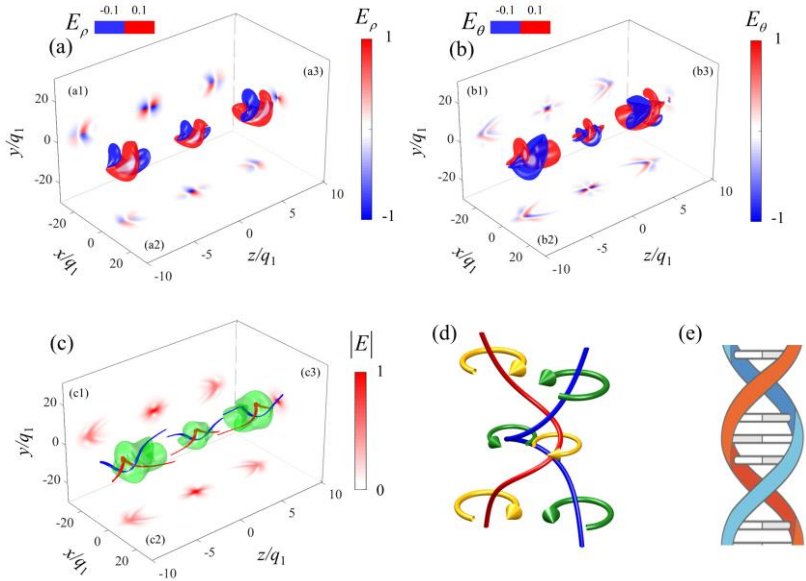

**Figure 1:** (a)-(c) displays the cross-sectional and contour topological images of $E_\rho$, $E_\theta$ and $|E|$ at time points $t=-7.5q_1/c$, $0$, $7.5q_1/c$, with parameters set to $q_1=0.01$ and $q_2=10q_1$. (a) The iso surface indicates the value of normalized intensity $E_\rho=\pm 0.1$. (a1/a2/a3) Section map of the $E_\rho$ components on the $y=0$ / $x=0$ / $z=0$ plane. (b) The iso surface indicates the value of normalized intensity $E_\theta=\pm 0.1$. (b1/b2/b3) Section map of the $E_\theta$ component on the $y=0$ / $x=0$ / $z=0$ plane. (c) The iso surface image of $|E|$, the iso surface is set to normalized intensity $|E|=0.1$. (c1/c2/c3) Section map of the $|E|$ component on the $y=0$ / $x=0$ / $z=0$ plane. In figure(c), the green iso surface indicates the value of normalized intensity $|E|=0.1$, and the red and blue curves are connected by the right-handed and left-handed vortex core, respectively. For clarity, we have only plotted the curve in the region where the wave packet is located, while the actual curve extends to infinity on both sides. All values in the graph have been normalized. (d) Schematic diagram of Double-helix Singularities. Red and blue represent the singularity lines, while yellow and green arrows indicate the direction of the electric field vectors. (e) The structure of DNA, adapted from Wikipedia

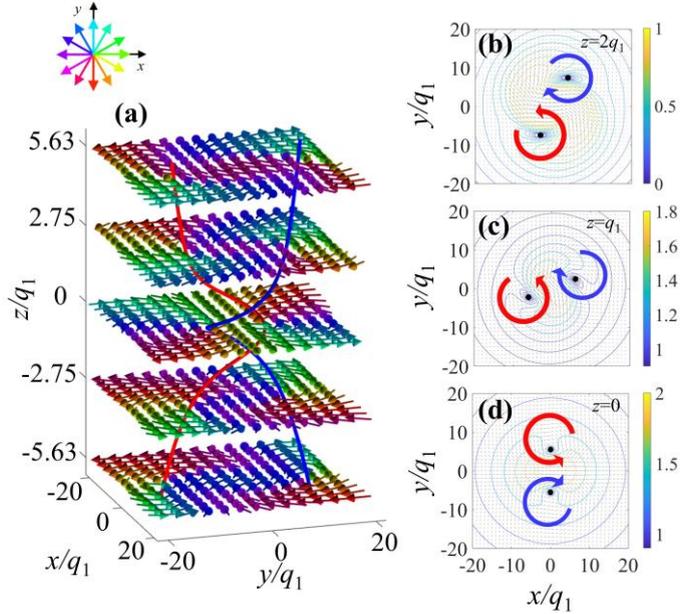

**Figure 2:** (a) shows electric field vector and singularity at the focusing time ( $t=0$ ). (b)-(d) shows the singularity lines and electric field vectors at the planes of $z=2q_1$, $q_1$, 0, and the red and blue arrows indicate the direction of the electric field. In (a), the color matching of electric field vector is related to its azimuth $\phi$ in the x-y plane, and the polar angle binds with brightness intensity, as shown in the illustration. The red and blue curves are respectively connected by the right-handed vortex core and the left-handed vortex core.

In the field of solid state physics, the phenomenon of paired generation and annihilation of vortices and antivortices is an important research topic [57, 58], as for the phenomenon of left-handed and right-handed vortices, it has also been observed in the field of Quantum fluid [59-61], Bose-Einstein condensate [62, 63]. However, in the field of optics, research focuses mainly on the generation and annihilation of vortex cores caused by the phase singularities with topology equal to 1 and -1 [64, 65]. To the best of the author's knowledge, this phenomenon has not yet been found in stable propagating topologic light quasi-particle. Our research has discovered this new phenomenon, opening up new directions for research in the field of complex electrodynamics. In addition, the phenomenon of stable left- handed and right-handed vortices intertwining with vortex core lines during propagation reminds us of the spiral vortex theory in fluid mechanics. The fluid particles rotate around the vortex line, and the velocity field forms two vortex fields around the vortex core, forming a structure of vortex dipoles. At the same time, the two vortex core lines spiral around each other. By considering the motion of electrons generated by TE electromagnetic pulses along the electric field, it can be demonstrated that there is a similarity between spiral vortices in the fluid and helical pulses. This similarity indicates that different physical phenomena may share topological features, revealing the similarities between linear and nonlinear systems in certain phenomena. On each transverse plane, the properties of vortex dipoles are similar to those of left-handed and right-handed vortices phase singularities. Under certain external conditions, the left-handed and right-handed vortices cores in the studied pulse can also approach each other and ultimately achieve mutual annihilation. This interesting phenomenon will become one of the important contents of future research. Given the topological protection properties of vortices, even in the face of small external disturbances, vortices and antivortices can still exist stably. This property will provide fundamental characteristic support for helical pulses in a wide range of large-scale applications.

## 4 Discussion

We discovered that helical pulses exhibit double-helix singularity and vortex-antivortex annihilation. This discovery not only enriches the content of spatiotemporal electromagnetic fields singularity families, but also provides a new perspective for the study of complex electrodynamics. Of particular note, we reported for the first time the dynamic processes of vortices and antivortices in structural optics, which reflects the driving role of interdisciplinary communication in research innovation. During the propagation process, the double helix structure of the singularity line of helical pulses remains topologically invariant. This characteristic, combined with the topological protection of the vortex, indicates that the controllable robustness of the singular structure has broad application prospects in the field of communication. At the same time, their unique electric field structure provides a research approach for studying transient phenomena in matter. Finally, we anticipate that this discovery will inspire potential applications in fields such as super-resolution microscopy, remote sensing, and lidar.

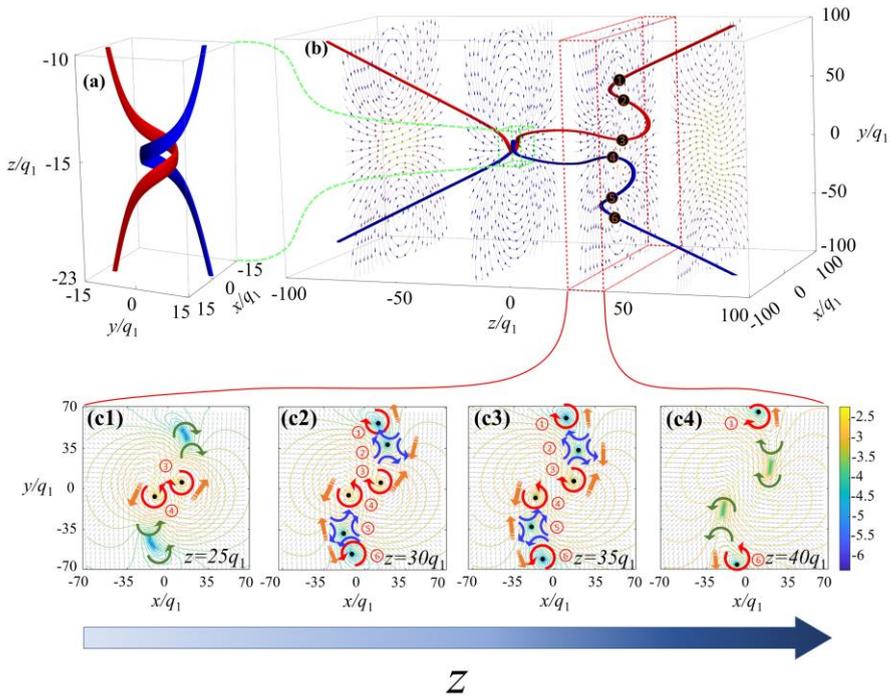

**Figure 3:** Electric field and singularity lines of helical pulses. (b) This figure presents the streamline plot of the electric field in the plane of the singular point and $z = -70q_1, -15q_1, 30q_1, 70q_1$ at the instant when $t = -15q_1/c$. (a) and (c) are enlarged local images of the red and green dashed rectangles in (b), respectively. The spiral structure maintains topological invariance. (c1-c4) displays the electric field vector and modulus contour lines on $z = 25q_1$, $z = 30q_1$, $z = 35q_1$ and $z = 40q_1$. In the image, the red curved arrow represents vortices, the blue curved arrow represents antivortices, the position of the green arrow indicates that vortices and antivortices are about to be generated or have already been annihilated, and the direction of the green arrow indicates the direction of the electric field vector. The orange dashed arrow indicates the direction of motion of the singularity. (c1-c4) reveal the variation of the singular points with different planes.

**Research funding:** This work has been supported by the National Natural Science Foundation of China (62171081, 61901086), the Natural Science Foundation of Sichuan Province (2022NSFSC0039), the Aeronautical Science Foundation of China (2023Z062080002), Singapore Ministry of Education (MOE) AcRF Tier 1 grant (RG157/23), MoE AcRF Tier 1 Thematic grant (RT11/23), Imperial-Nanyang Technological University Collaboration Fund (INCF-2024-007), and a Start Up Grant of Nanyang Technological University.

**Author contribution:** R.W. conceived the ideas and supervised the project, S.S. and M.X. performed the theoretical modeling and numerical simulations, R.W., Y.S., and S.S. conducted data analysis. All authors wrote the manuscript and participated the discussions. All authors have accepted responsibility for the entire content of this manuscript and approved its submission.

**Conflict of interest:** Authors state no conflicts of interest.

**Ethical approval:** The conducted research is not related to either human or animals use.

**Data availability statement:** Data generated in this study is available from the corresponding authors upon reasonable request.

## References

[1] S. Lindsay, T. Thundat, L. Nagahara, U. Knipping, and R. Rill, "Images of the DNA double helix in water," *Science*, vol. 244, no. 4908, pp. 1063-1064, 1989.

[2] J. I. Bell, "The double helix in clinical practice," *Nature*, vol. 421, no. 6921, pp. 414-416, 2003.

[3] R. S. Mathew-Fenn, R. Das, and P. A. Harbury, "Remeasuring the double helix," *Science*, vol. 322, no. 5900, pp. 446-449, 2008.

[4] S. Akamatsu, M. Perrut, S. Bottin-Rousseau, and G. Faivre, "Spiral Two-Phase Dendrites," *Physical Review Letters*, vol. 104, no. 5, 2010, Art no. 056101.

[5] M. J. Williams and M. Bachmann, "Stabilization of helical macromolecular phases by confined bending," *Physical review letters*, vol. 115, no. 4, 2015, Art no. 048301.

[6] X.-Q. Sun, B. Lian, and S.-C. Zhang, "Double helix nodal line superconductor," *Physical review letters*, vol. 119, no. 14, 2017, Art no. 147001.

[7] Y. Li, M. Zhou, Y. Song, T. Higaki, H. Wang, and R. Jin, "Double-helical assembly of heterodimeric nanoclusters into supercrystals," *Nature*, vol. 594, no. 7863, pp. 380-384, 2021.

[8] C. Donnelly et al., "Complex free-space magnetic field textures induced by three-dimensional magnetic nanostructures," *Nature nanotechnology*, vol. 17, no. 2, pp. 136-142, 2022.

[9] H. Zhang, J. Zeng, X. Y. Lu, Z. Y. Wang, C. L. Zhao, and Y. J. Cai, "Review on fractional vortex beam," *Nanophotonics*, vol. 11, no. 2, pp. 241-273, 2022.

[10] Y. Shen et al., "Optical vortices 30 years on: OAM manipulation from topological charge to multiple singularities," *Light: Science & Applications*, vol. 8, no. 1, 2019, Art no. 90.

[11] J. Wang, J. Liu, S. H. Li, Y. F. Zhao, J. Du, and L. Zhu, "Orbital angular momentum and beyond in free-space optical communications," *Nanophotonics*, vol. 11, no. 4, pp. 645-680, 2022.

[12] J. W. Wang et al., "Experimental observation of Berry phases in optical Mobius-strip microcavities," *Nature Photonics*, vol. 17, no. 1, pp. 120-+, 2023.

[13] Y. Song et al., "Mobius Strip Microlasers: A Testbed for Non-Euclidean Photonics," *Physical Review Letters*, vol. 127, no. 20, 2021, Art no. 203901.

[14] Y. Li, M. A. Ansari, H. Ahmed, R. X. Wang, G. C. Wang, and X. Z. Chen, "Longitudinally variable 3D optical polarization structures," *Science Advances*, vol. 9, no. 47, 2023, Art no. eadj6675.

[15] H. Larocque, A. D'Errico, M. F. Ferrer-Garcia, A. Carmi, E. Cohen, and E. Karimi, "Optical framed knots as information carriers," *Nature Communications*, vol. 11, no. 1, 2020, Art no. 5119.

[16] Y. N. Zhang, J. Gao, F. Xia, B. Han, and Y. Zhao, "Microfiber Knot Resonators: Structure, Spectral Properties, and Sensing Applications," *Laser Photon. Rev.*, vol. 18, no. 1, Jan 2024, Art no. 24.

[17] Y. Shen, Z. Wang, X. Fu, D. Naidoo, and A. Forbes, "SU(2) Poincar\'e sphere: A generalized representation for multidimensional structured light," *Physical Review A*, vol. 102, no. 3, 2020, Art no. 031501.

[18] M. Z. Liu et al., "Broadband generation of perfect Poincare beams via dielectric spin-multiplexed metasurface," *Nature Communications*, vol. 12, no. 1, 2021, Art no. 2230.

[19] C. P. Jisha, S. Nolte, and A. Alberucci, "Geometric Phase in Optics: From Wavefront Manipulation to Waveguiding," *Laser Photon. Rev.*, vol. 15, no. 10, 2021, Art no. 2100003.

[20] Y. Shen, E. C. Martínez, and C. Rosales-Guzmán, "Generation of Optical Skyrmions with Tunable Topological Textures," *ACS Photonics*, vol. 9, no. 1, pp. 296-303, 2022.

[21] Y. Shen, Q. Zhang, P. Shi, L. Du, X. Yuan, and A. V. Zayats, "Optical skyrmions and other topological quasiparticles of light," *Nature Photonics*, vol. 18, no. 1, pp. 15-25, 2024.

[22] H. O. Teng, J. I. Zhong, J. I. Chen, X. R. Lei, and Q. I. Zhan, "Physical conversion and superposition of optical skyrmion topologies," *Photonics Res.*, vol. 11, no. 12, pp. 2042-2053, 2023.

[23] Y. Shen et al., "Topologically controlled multiskyrmions in photonic gradient-index lenses," *Physical Review Applied*, vol. 21, no. 2, 2024, Art no. 024025.

[24] M. Lin, Q. Liu, H. G. Duan, L. P. Du, and X. C. Yuan, "Wavelength-tuned transformation between photonic skyrmion and meron spin textures," *Applied Physics Reviews*, vol. 11, no. 2, 2024, Art no. 021408.

[25] C. Wan, Y. Shen, A. Chong, and Q. Zhan, "Scalar optical hopfions," *eLight*, vol. 2, no. 1, pp. 1-7, 2022.

[26] Y. Shen, B. Yu, H. Wu, C. Li, Z. Zhu, and A. V. Zayats, "Topological transformation and free-space transport of photonic hopfions," *Advanced Photonics*, vol. 5, no. 1, 2023, Art no. 015001.

[27] D. Sugic et al., "Particle-like topologies in light," *Nature Communications*, vol. 12, no. 1, 2021, Art no. 6785.

[28] Y. Shen, X. Fu, and M. Gong, "Truncated triangular diffraction lattices and orbital-angular-momentum detection of vortex SU(2) geometric modes," *Optics Express*, vol. 26, no. 20, pp. 25545-25557, 2018.

[29] G. J. Gbur, *Singular optics*. New York, USA: CRC press, 2016.

[30] H. Ahmed et al., "Optical metasurfaces for generating and manipulating optical vortex beams," *Nanophotonics*, vol. 11, no. 5, pp. 941-956, 2022.

[31] J. Peng, R.-Y. Zhang, S. Jia, W. Liu, and S. Wang, "Topological near fields generated by topological structures," *Science Advances*, vol. 8, no. 41, 2022, Art no. eabq0910.

[32] Z. Y. Che et al., "Polarization Singularities of Photonic Quasicrystals in Momentum Space," *Physical Review Letters*, vol. 127, no. 4, 2021, Art no. 043901.

[33] G. Arora, S. Joshi, H. Singh, V. Haridas, and P. Senthilkumaran, "Perturbation of V-point polarization singular vector beams," *Opt. Laser Technol.*, vol. 158, 2023, Art no. 108842.

[34] Y. Shen, X. Yang, D. Naidoo, X. Fu, and A. Forbes, "Structured ray-wave vector vortex beams in multiple degrees of freedom from a laser," *Optica*, vol. 7, no. 7, pp. 820-831, 2020.

[35] D. Mao et al., "Generation of polarization and phase singular beams in fibers and fiber lasers," *Advanced Photonics*, vol. 3, no. 1, 2021, Art no. 014002.

[36] G. F. Q. Rosen, P. I. Tamborenea, and T. Kuhn, "Interplay between optical vortices and condensed matter," *Rev. Mod. Phys.*, vol. 94, no. 3, 2022, Art no. 035003.

[37] E. Prinz, M. Hartelt, G. Spektor, M. Orenstein, and M. Aeschlimann, "Orbital Angular Momentum in Nanoplasmonic Vortices," *ACS Photonics*, vol. 10, no. 2, pp. 340-367, 2023.

[38] H.-J. Wu et al., "Conformal frequency conversion for arbitrary vectorial structured light," *Optica*, vol. 9, no. 2, pp. 187-196, 2022.

[39] P. Ornelas, I. Nape, R. D. Koch, and A. Forbes, "Non-local skyrmions as topologically resilient quantum entangled states of light," *Nature Photonics*, vol. 18, no. 3, 2024, Art no. 10.

[40] H. Ren and S. A. Maier, "Nanophotonic Materials for Twisted-Light Manipulation," *Adv. Mater.*, vol. 35, no. 34, 2023, Art no. 2106692.

[41] Q. C. Yan et al., "Quantum Topological Photonics," *Adv. Opt. Mater.*, vol. 9, no. 15, 2021, Art no. 2001739.

[42] Z. Wan, Y. Shen, Z. Wang, Z. Shi, Q. Liu, and X. Fu, "Divergence-degenerate spatial multiplexing towards future ultrahigh capacity, low error-rate optical communications," *Light: Science & Applications*, vol. 11, no. 1, 2022, Art no. 144.

[43] A. Pryamikov, "Rising complexity of the OAM beam structure as a way to a higher data capacity," *Light: Science & Applications*, vol. 11, no. 1, 2022, Art no. 221.

[44] Z. Wan, H. Wang, Q. Liu, X. Fu, and Y. Shen, "Ultra-Degree-of-Freedom Structured Light for Ultracapacity Information Carriers," *ACS Photonics*, vol. 10, no. 7, pp. 2149-2164, 2023.

[45] Y. J. Shen et al., "Roadmap on spatiotemporal light fields," *Journal of Optics*, vol. 25, no. 9, 2023, Art no. 093001.

[46] R. Hellwarth and P. Nouchi, "Focused one-cycle electromagnetic pulses," *Physical Review E*, vol. 54, no. 1, pp. 889-895, 1996.

[47] A. Zdagkas et al., "Observation of toroidal pulses of light," *Nature Photonics*, vol. 16, no. 7, pp. 523-528, 2022.

[48] K. Jana et al., "Quantum control of flying doughnut terahertz pulses," *Science Advances*, vol. 10, no. 2, p. 7, 2024.

[49] R. Wang et al., "Observation of resilient propagation and free-space skyrmions in toroidal electromagnetic pulses," *Applied Physics Reviews*, vol. 11, no. 3, 2024.

[50] Y. Shen, A. Zdagkas, N. Papasimakis, and N. I. Zheludev, "Measures of space-time nonseparability of electromagnetic pulses," *Physical Review Research*, vol. 3, no. 1, 2021, Art no. 013236.

[51] Y. Shen, Y. Hou, N. Papasimakis, and N. I. Zheludev, "Supertoroidal light pulses as electromagnetic skyrmions propagating in free space," *Nature communications*, vol. 12, no. 1, 2021, Art no. 5891.

[52] T. Raybould, V. A. Fedotov, N. Papasimakis, I. Youngs, and N. I. Zheludev, "Exciting dynamic anapoles with electromagnetic doughnut pulses," *Applied Physics Letters,* vol. 111, no. 8, 2017, Art no. 081104.

[53] Y. Shen, N. Papasimakis, and N. I. Zheludev, "Nondiffracting supertoroidal pulses and optical "Kármán vortex streets"," *Nature Communications*, vol. 15, no. 1, 2024, Art no. 4863.

[54] A. Zdagkas, N. Papasimakis, V. Savinov, M. R. Dennis, and N. I. Zheludev, "Singularities in the flying electromagnetic doughnuts," *Nanophotonics*, vol. 8, no. 8, pp. 1379-1385, 2019.

[55] J. Lekner, "Localized electromagnetic pulses with azimuthal dependence," *Journal of Optics A: Pure and Applied Optics*, vol. 6, no. 7, pp. 711-716, 2004.

[56] R. W. Ziolkowski, "Localized transmission of electromagnetic energy," *Physical Review A*, vol. 39, no. 4, pp. 2005-2033, 1989.

[57] F. P. Chmiel et al., "Observation of magnetic vortex pairs at room temperature in a planar α-Fe2O3/Co heterostructure," *Nature Materials*, vol. 17, no. 7, pp. 581-585, 2018.

[58] Y.-J. Wang et al., "Polar Bloch points in strained ferroelectric films," *Nature Communications*, vol. 15, no. 1, 2024, Art no. 3949.

[59] G. Gauthier et al., "Giant vortex clusters in a two-dimensional quantum fluid," *Science*, vol. 364, no. 6447, pp. 1264-+, 2019.

[60] W. J. Kwon et al., "Sound emission and annihilations in a programmable quantum vortex collider," *Nature*, vol. 600, no. 7887, pp. 64-69, 2021.

[61] T. Congy, P. Azam, R. Kaiser, and N. Pavloff, "Topological Constraints on the Dynamics of Vortex Formation in a Two-Dimensional Quantum Fluid," *Physical Review Letters*, vol. 132, no. 3, 2024, Art no. 033804.

[62] T. Kanai and W. Guo, "True mechanism of spontaneous order from turbulence in two-dimensional superfluid manifolds," *Physical Review Letters*, vol. 127, no. 9, 2021, Art no. 095301.

[63] A. J. Groszek, M. J. Davis, and T. P. Simula, "Decaying quantum turbulence in a two-dimensional Bose-Einstein condensate at finite temperature," *SciPost Phys.*, Article vol. 8, no. 3, 2020, Art no. 039.

[64] H. L. Lin et al., "Optical vortex-antivortex crystallization in free space," *Nature Communications*, vol. 15, no. 1, 2024, Art no. 6178.

[65] H. L. Lin, S. H. Fu, H. Yin, Z. Li, and Z. Q. Chen, "Intrinsic Vortex-Antivortex Interaction of Light," *Laser Photon. Rev.*, vol. 16, no. 8, 2022, Art no. 2100648.